\documentclass[12pt]{article}

\usepackage[utf8]{inputenc}
\usepackage[english]{babel}
\usepackage{amsmath, amsfonts, amssymb, amsthm}
\usepackage{mathtools}
\usepackage{float}
\mathtoolsset{showonlyrefs}
\usepackage{geometry}
\usepackage{graphicx}
\usepackage{natbib} 
\usepackage{xr-hyper}
\usepackage{hyperref}
\usepackage{setspace} 
\usepackage{lineno}   
\usepackage{indentfirst}
\usepackage{xcolor}
\usepackage[most]{tcolorbox}
\usepackage{algorithm}
\usepackage{algpseudocode}
\usepackage{array,tabularx,booktabs}
\usepackage{longtable}
\usepackage{url}
\newcolumntype{Y}{>{\centering\arraybackslash}X}
\newcommand{\yc}[1]{{{\color{red}  #1}}}

\newcommand{\rv}{\textcolor{black}}

\theoremstyle{definition}
\newtheorem{exmp}{Example}

\newenvironment{examplecont}
{\addtocounter{exmp}{-1} 
 \begin{exmp}[continued]}
{\end{exmp}}

\newtheorem{theorem}{Theorem}

\newtheorem{assumption}{Assumption}

\title{\Large Sparse Latent Class Analysis \rv{for Dichotomous Responses}: Post-Estimation Refinement via Item-Level Pseudo-Likelihood}

\author{
Yuxuan Xu$^{1}$,
Lea Kaufmann$^{2}$,
Yunxiao Chen$^{1,*}$,\\
Maria Kateri$^{2}$,
and Irini Moustaki$^{1}$
\\[0.6em]
\small $^{1}$Department of Statistics,
London School of Economics and Political Science, London, UK
\\
\small $^{2}$Institute of Statistics,
RWTH Aachen University, Aachen, Germany
}

\date{} 

\begin{document}

\maketitle
\vspace{-1cm}
\begin{abstract}
Latent Class Analysis (LCA) is widely used to identify unobserved subgroups in social and behavioural sciences. A long-standing challenge for LCA is the interpretability of the latent classes, due to the high complexity of the estimated item response probability matrix.
To address this, we propose a computationally efficient post-estimation refinement procedure that enhances model interpretability by a sparse model estimate. The method begins by estimating a classical, unrestricted, latent class model and determining the number of classes using the Bayesian information criterion (BIC).
It is followed by a refinement step that further performs model selection on the item-specific response probabilities based on the initial estimate. This refinement penalises the number of distinct response probability levels per item, collapsing redundant levels to yield a sparse matrix that is significantly easier to interpret than those produced by classical LCA. We provide asymptotic theory showing that the proposed procedure consistently recovers the sparse pattern of the item response probabilities for each item, and further validate its performance through extensive simulations. The practical power of the proposed method is further illustrated via an application to survey data on social role performance, where it provides a parsimonious and clear characterisation of the resulting latent classes. The code for implementing the proposed method is publicly available at \url{https://github.com/florence07/Sparse-LCA-Refinement}.
\end{abstract}



\newpage

\section{Introduction}
Rooted in the seminal contributions of \cite{lazarsfeld1968latent} and \cite{goodman1974a,goodman1974b}, Latent Class Analysis (LCA) has emerged as a cornerstone for exploratory data analysis, enabling researchers to identify and characterise unobserved subgroups within heterogeneous populations. This versatile framework has found widespread applications across the social and behavioural sciences, public health, and marketing research; for a rigorous treatment of its inferential foundations and broad empirical utility, we refer the reader to \cite{collins2010latent}. Beyond these roots in exploratory data analysis, latent class models are increasingly employed to assess fine-grained psychological and educational attributes, a field commonly referred to as cognitive diagnosis or diagnostic classification (see \cite{Ruppetal2010} and \cite{vondavier2019handbook} for comprehensive reviews).
These models typically impose restrictions on the parameter space of standard latent class models and are therefore often known as restricted latent class models \citep{gu2020partial}. 
Notwithstanding these measurement-focused developments, the current paper remains situated within the exploratory tradition of LCA. 

Assigning substantive meaning to latent classes is central to exploratory LCA. This step is typically done by examining the item-response probability matrix. These parameters essentially define the `profile' of each subgroup; yet, their interpretation is often non-trivial. With many latent classes or items, identifying distinct, interpretable patterns within the matrix becomes difficult. This often yields `noisy' classes that are difficult to distinguish, creating a risk that subgroup labelling becomes inaccurate or overly subjective. To address interpretability challenges, recent advances have focused on inducing sparsity in the item response matrix.  \cite{chen2017regularized} proposed a regularised estimation approach for the LCA of binary response data. This approach induces sparsity in the item-response probability matrix, thereby improving the interpretability of the resulting latent classes.
This approach has been further extended to polytomous response data in \cite{robitzsch2020regularized}. However, this regularised estimation approach requires maximising a regularised log-likelihood function that includes a nonconvex, nonsmooth penalty term. The optimisation is typically carried out via an Expectation-Maximisation (EM) algorithm, for which each M-step requires solving multiple nonconvex, nonsmooth optimisation problems. Such a heavy computational burden may render the method impractical for many real-world applications. Furthermore, within exploratory data analysis settings, sparsity assumptions and regularised estimation procedures have been applied to diagnostic classification models to enhance the interpretability of latent classes (see, e.g., \cite{chen2015statistical}, \cite{chen2020sparse}, \cite{wang2021learning}, and \cite{he2025sparse}). \rv{Recent work has also established identification results for this type of sparse structure. \cite{bowers2024ESRLCM} proposed the Equivalence Set Restricted Latent Class Model (ESRLCM), which induces the sparse structure in the item-response probability matrix through equivalence-set restrictions, and established sufficient conditions for generic identifiability based on whether the item-response structure can adequately distinguish the latent classes.}

In this paper, we propose a post-estimation refinement procedure that achieves the interpretative benefits of sparsity without the prohibitive computational costs of regularised estimation. Our approach operates in two stages: we first estimate a classical LCA model and determine the number of classes via the Bayesian Information Criterion (BIC; \citealp{schwarz1978estimating}), and then apply a refinement step 
that yields a sparse item-response probability matrix with clearer patterns, thereby making the latent classes 
significantly easier to interpret.  
This refinement procedure has two building blocks: (1) a stepwise search algorithm for exploring sparse models and (2) an Extended-BIC (EBIC) for comparing models with different sparsity levels,  both of which rely on a novel item-specific pseudo-likelihood function constructed from the initial model estimate.  
We provide asymptotic theory confirming that our procedure consistently selects the number of distinct item-response probability levels. Furthermore, we demonstrate the power of this method through extensive simulations and an application to survey data on social role performance, where it clarifies respondents' complex profiles. 

The item-specific pseudo-likelihood function mentioned above plays a central role in the proposed refinement procedure. This function is constructed from the posterior probabilities of observations' latent class membership, computed under the initial model estimate in the first stage. Since the initial model estimate is already root-$N$ consistent \citep{collins2010latent}, these posterior probabilities are sufficiently close to those we would obtain under the true model. Consequently, the resulting EBIC enjoys similar asymptotic properties as the classical BIC and thus yields statistical consistency in model selection. 
Similar pseudo-likelihood functions have been used in stepwise estimation methods for latent class models, such as the ones developed to model a distal outcome predicted by a latent categorical variable \citep{zhu2017general}. However, existing work focuses on parameter estimation, whereas, to our best knowledge, the proposed method is the first to use such a pseudo-likelihood for model selection. 
Further discussions about stepwise methods for latent class models can be found in \cite{zhu2017general},  \cite{bakk2018two}, and \cite{bakk2021relating}.

The rest of the paper is organised as follows. Section 2 introduces the proposed refinement method and establishes its theoretical guarantees.  Section 3 presents a comprehensive simulation study to evaluate the method’s performance. Section 4 illustrates the practical utility of the approach by applying it to data from a health survey. Finally, Section 5 presents a discussion of the results and concludes.

\section{The Proposed Method}

\subsection{Latent Class Analysis}
We consider $N$ respondents answering $J$ items. Let $Y_{ij}$ denote the response from respondent $i$ to item $j$. To simplify the discussion, we assume that the responses are binary, i.e., $Y_{ij} \in \{0, 1\}$. As discussed in Section \ref{sec:discussion}, the proposed method can be extended to data with polytomous items.  

Suppose the data follow a latent class model with $K$ classes. Let $\xi_i \in \{1, 2, ..., K\}$ indicate the latent class membership of respondent $i$. The model assumes that $\xi_i$, $i = 1, ..., N$, are independent and identically distributed (i.i.d.) random variables, following a categorical distribution, with
$P(\xi_i = k) = \nu_k$, where $\nu_1$, ..., $\nu_K$ are unknown parameters satisfying $\nu_k \geq 0$ and $\nu_1 + \cdots + \nu_K = 1$. Moreover, the model assumes that an individual's response probabilities depend only on their latent class membership. That is, 
$$P(Y_{ij} = 1\vert \xi_i = k) = \beta_{jk},$$
where $\beta_{jk}$, $j = 1, ..., J$ and $k=1, ..., K$, are unknown parameters. Finally, the local independence assumption is adopted, which says that for each respondent $i$, $Y_{i1}$, ..., $Y_{iJ}$ are conditionally independent given $\xi_i$. 

For the parsimony of notation, we use $\boldsymbol{\nu} = (\nu_1, ..., \nu_K)^\top$, $\mathbf B = (\beta_{jk})_{J\times K}$, and $\boldsymbol{\beta}_j = (\beta_{j1}, ..., \beta_{jK})^\top$ to denote the parameter vectors or matrix, and 
$\boldsymbol{\nu}^* = (\nu_1^*, ..., \nu_K^*)^\top$, $\mathbf B^* = (\beta_{jk}^*)_{J\times K}$, and $\boldsymbol{\beta}_j^* = (\beta_{j1}^*, ..., \beta_{jK}^*)^\top$ the corresponding true values. 
Under the assumptions of the latent class model, the marginal log-likelihood function takes the form 
\begin{equation}
l(\boldsymbol{\nu}, \mathbf B) = \sum_{i=1}^N\log\Big(\sum_{k=1}^K \nu_k\prod_{j=1}^J \beta_{jk}^{Y_{ij}}(1-\beta_{jk})^{1-Y_{ij}}\Big).    
\end{equation}
Classical LCA estimates the unknown model parameters by the maximum likelihood estimator (MLE), i.e., 
\begin{equation}\label{eq:MLE}
(\hat{\boldsymbol{\nu}}, \hat{\mathbf B}) = \arg\max_{(\boldsymbol{\nu}, \mathbf B)} ~l(\boldsymbol{\nu}, \mathbf B).
\end{equation}
This optimisation is typically solved using an Expectation-Maximisation (EM) algorithm. 
Under correct model specification and mild regularity conditions, the MLE is root-$N$ consistent, in the sense that 
\begin{equation}
\Vert \hat{\boldsymbol{\nu}} - {\boldsymbol{\nu}}^*\Vert = O_p(1/\sqrt{N}) \mbox{~and~} \Vert \hat{\mathbf B} - {\mathbf B}^* \Vert_F = O_p(1/\sqrt{N}),    
\end{equation}
where $\Vert\cdot\Vert$ denotes the Euclidean norm and $\Vert\cdot \Vert_F$ denotes the matrix Frobenius norm\footnote{There is a label indeterminacy issue, also known as the label switching problem, that is related to this consistency result. Here, we assume that the true and estimated labels have been consistently aligned, which can be achieved, for example, by assuming $\nu_1^* >  \cdots > \nu_K^*$ and imposing the same ordering on $\hat \nu_1$, ..., $\hat \nu_K$.}. 

When the number of latent classes $K$ is unknown, the optimal number of latent classes is chosen by fitting models with increasing numbers of classes and selecting the one that minimises the BIC value. 
We refer readers to Chapter 4, \cite{collins2010latent}, for a review of the inferential foundations of the classical LCA. For ease of exposition, we treat the number of classes $K$ as known when motivating and describing the proposed refinement method in Sections~\ref{subsec:slca} and \ref{subsec:refinement} below.

\subsection{Sparse LCA} \label{subsec:slca}

Interpreting the latent classes is an essential step for LCA. This step is based on $\hat{\mathbf B}$, which characterises the item response behaviours of all the learned latent classes. However, when $J$ or $K$ is moderately large, interpreting the latent classes can be a challenge, due to the large number of parameters in $\hat B$. To tackle this challenge, \cite{chen2017regularized} proposed a sparse latent class model, which assumes that the number of distinct item response probability levels for each item can be smaller than the total number of classes. 

More specifically, let $m_j^*  =\text{card}(\boldsymbol{\beta}^*_{j})$ denote the true number of distinct item response probability levels for item $j$. The sparse latent class model assumes that $m_j^* < K$ for many items.
However, even when the true model is sparse, due to the randomness in data, each $\hat{\boldsymbol{\beta}}_{j}$ has $K$ distinct values with probability tending to 1 as the sample size goes to infinity, and thus, cannot recover the sparse pattern in the true model. 
To tackle this issue, \cite{chen2017regularized} proposed a regularised estimation method that simultaneously learns the number of distinct 
item response probability levels, denoted by $\tilde m_j$, and estimates the item response probabilities, denoted by $\bar{\mathbf B} = (\bar \beta_{jk})_{J\times K}$, satisfying that the number of distinct levels in $(\bar \beta_{j1}, ..., \bar \beta_{jK})^\top$ is $\tilde m_j$, where $\tilde m_j$ can be strictly smaller than $K$. Consequently, $\bar{\mathbf B}$ tends to have substantially fewer parameters than those in $\hat {\mathbf B}$, the MLE for classical LCA, and thus, tends to be much easier to interpret. Consistency for both the model selection and parameter estimation has been established for this regularised estimator.

Despite the interpretational advantage and theoretical guarantees, this regularised estimator requires maximising a regularised log-likelihood function, which involves complex Expectation-Maximisation (EM) cycles in which each M-step involves solving multiple non-smooth optimisation problems. Such a heavy computational burden renders the method impractical for many real-world applications.  

\begin{exmp}\label{exmp:1}
Table~\ref{tab:example} gives a toy example of the true item response probability matrix $\mathbf B$ for a sparse latent class model, with $J= 6$ items and $K=4$ latent classes.  In this example, the elements of $\boldsymbol{\beta}_j$ have two distinct levels for items 1 to 5, and three distinct levels for item 6. If such a sparse pattern can be recovered from the data, one can infer that the first latent class has the lowest item response probabilities across all items, while the fourth latent class has the highest item response probabilities. In addition, each of the second and third classes has a combination of the lowest and highest item response probabilities for items 1 to 5, and a median level of item response probability for item 6. In a classical latent class model, however, each item would typically be assigned four distinct response-probability levels, even when the true underlying structure is sparse, making the resulting classes harder to interpret.

\begin{table}[h]
    \centering
    \begin{tabular}{c|cccc}
    \hline
        &  \multicolumn{4}{c}{Latent Class}\\
       Item & 1 & 2 & 3 & 4\\
       \hline
       1& 0.10 & 0.10 & 0.10 & 0.90\\
       2& 0.20 & 0.20 & 0.80 & 0.80\\
       3& 0.15 & 0.75 & 0.15 & 0.75\\
       4& 0.30 & 0.30 & 0.30 & 0.85 \\
       5& 0.25 & 0.25 & 0.90 & 0.90\\
       6& 0.20 & 0.55 & 0.55 & 0.95\\
    \hline
    \end{tabular}
    \caption{An example of the item response probabilities for a sparse latent class model.}
    \label{tab:example}
\end{table}
\end{exmp}

\subsection{Proposed Refinement Procedure}\label{subsec:refinement}

We propose a refinement procedure to estimate a sparse latent class model, which remains statistically consistent while avoiding the high computational burden of the regularised estimator. Like 
the regularisation approach in \cite{chen2017regularized}, our goal is to find an estimate of the latent class model, denoted by $(\bar{\boldsymbol\nu}, \bar{\mathbf B})$, such that $\bar{\mathbf B}$ is sparse, in the sense that each row of $\bar{\mathbf B}$ has a small number of distinct values, i.e., each item has a small number of unique item response probability levels.

The proposed estimator has two building blocks, including (1) a stepwise search algorithm for exploring sparse models and  (2) an EBIC for comparing models with different sparsity levels, both of which rely on an item-specific pseudo-likelihood function constructed based on the initial model estimate from the first stage. In what follows, we first define the pseudo-likelihood, then describe the stepwise search algorithm, and finally the EBIC. 

\medskip
\noindent
{\bf Item-specific pseudo-likelihood.} 
Let 
\begin{equation}
\hat \gamma_{ik} = \frac{\hat\nu_k\prod_{j=1}^J \hat\beta_{jk}^{Y_{ij}}(1-\hat\beta_{jk})^{1-Y_{ij}}}{\sum_{h=1}^K\hat\nu_h\prod_{j=1}^J \hat\beta_{jh}^{Y_{ij}}(1-\hat\beta_{jh})^{1-Y_{ij}}}
\end{equation}
be the posterior probability of respondent $i$ belonging to the $k$th latent class evaluated under the MLE for the classical LCA, and we denote $\hat{\mathbf{\Gamma}} = \{\hat{\gamma}_{ik}\}_{N\times K}$. The pseudo-likelihood function for item $j$ is defined as 
\begin{equation}\label{eq:pseudo-lik}
Q_j(\boldsymbol{\beta}_j ; \hat{\mathbf{\Gamma}}) = \sum_{i=1}^N  \sum_{k=1}^K \hat \gamma_{ik} \left( Y_{ij} \log(\beta_{jk}) + (1 - Y_{ij}) \log(1 - \beta_{jk}) \right).    
\end{equation}
This is a weighted Bernoulli log-likelihood with the latent class memberships replaced by the corresponding posterior probabilities $\hat{\gamma}_{ik}$; for brevity, we refer to it as the pseudo-likelihood throughout. We should also note that this function is the objective function for $\boldsymbol{\beta}_j$ in the M-step of the EM algorithm for solving \eqref{eq:MLE}.

This item-specific pseudo-likelihood enables independent model selection item by item, substantially reducing the complexity of the model selection problem.

\medskip
\noindent
{\bf Stepwise search algorithm.} For each item $j$ and each possible number of distinct item response probability levels $m_j \in \{1, \dots, K\}$, ideally, we hope to solve the following maximisation problem 
\begin{equation}\label{eq:cmle}
\tilde{\boldsymbol{\beta}}_{j}^{(m_j)} = \arg\max_{\boldsymbol{\beta}_{j}} ~Q_j(\boldsymbol{\beta}_{j} ; \hat{\mathbf{\Gamma}}), \quad \text{s.t. } \text{card}(\boldsymbol{\beta}_{j}) = m_j.
\end{equation}
Recall that $\text{card}(\boldsymbol{\beta}_{j})$ denotes the number of distinct values within the $j$-th item response probability vector $\boldsymbol{\beta}_{j}$. In other words, 
$\tilde{\boldsymbol{\beta}}_{j}^{(m_j)}$  reestimates the response probabilities for item $j$, under the constraint that the item has $m_j$ distinct item response probability levels. The solutions 
$\tilde{\boldsymbol{\beta}}_{j}^{(1)}$, ..., $\tilde{\boldsymbol{\beta}}_{j}^{(K)}$
are then compared using the EBIC introduced below. Unfortunately, \eqref{eq:cmle} becomes computationally infeasible when $K$ is sufficiently large, due to the combinatorial nature of the discrete constraint $\text{card}(\boldsymbol{\beta}_{j}) = m_j$.  For example, when $K = 10$, solving \eqref{eq:cmle} for $m_j = 1, ..., 10$ requires comparing a total of 115,975 models, which is computationally demanding. 

To tackle this computational challenge, we instead use a stepwise search algorithm that leverages the ordering of the initial estimate $\hat{\boldsymbol\beta}_j$. 
The algorithm
iteratively identifies partitions of the entries of $\boldsymbol{\beta}_j$ such that the parameters in the same subset of the partition share the same value. More specifically, let $\mathcal C = \{C_1, ..., C_{m_j}\}$ be a partition of $\{1, ..., K\}$, satisfying that (1) each subset $C_l$ is nonempty, (2) $C_l$s are mutually exclusive, and (3) the union of $C_l$ is $\{1, ..., K\}$. 
A partition defines a constrained parameter space for $\boldsymbol{\beta}_j$, 
\[
\Theta_j(\mathcal C)
=
\{\boldsymbol{\beta}_j:\beta_{jk}=\beta_{jk'}, \forall\, k,k'\in C_l,\ l=1, ..., m_j, \mbox{and~} \beta_{jk} <  \beta_{jk'}, \mbox{~if~} k\in C_l, k' \in C_{l'}, l < l'\},
\]
satisfying $\text{card}(\boldsymbol{\beta}_{j}) = m_j$ for all  
 $\boldsymbol{\beta}_{j} \in \Theta_j(\mathcal C)$.  As $\hat{\boldsymbol\beta}_j$ 
is a statistically consistent estimate of ${\boldsymbol\beta}_j^*$, the ordering of the estimated parameters in $\hat{\boldsymbol\beta}_j$ is useful when we search for the partition based on the true parameters $\boldsymbol{\beta}_j^*$. 

For ease of exposition, we assume that $\hat{\boldsymbol\beta}_j$ has $K$ distinct entries, which is typically the case in practice and asymptotically guaranteed. Let $\pi_j$ be a permutation of $\{1,\ldots,K\}$
that defines the order statistic for $\hat \beta_{j1}$, ..., $\hat \beta_{jK}$, satisfying that $\hat\beta_{j,\pi_j(1)} <  \cdots < \hat\beta_{j,\pi_j(K)}$. The algorithm starts with $\tilde{\boldsymbol{\beta}}_{j}^{(K)} = \hat{\boldsymbol\beta}_j$. This solution defines a partition 
$\mathcal C_j^{(K)}= \bigl\{\{\pi_j(1)\},\ldots,\{\pi_j(K)\}\bigr\}$, where each subset contains exactly one latent class and the subsets are ordered based on the initial estimate $\hat{\boldsymbol{\beta}}_j$. It is easy to see that 
\[\tilde{\boldsymbol{\beta}}_{j}^{(K)} = \arg\max_{\boldsymbol{\beta}_j}
        Q_j(\boldsymbol{\beta}_{j};\hat{\mathbf{\Gamma}}) 
        \quad \text{s.t. } \boldsymbol{\beta}_{j} \in \Theta_j(\mathcal C_j^{(K)}).\]
The algorithm then proceeds to find a partition 
$\mathcal C_j^{(K-1)}$ by merging two subsets of $\mathcal C_j^{(K)}$, and then learns 
\[\tilde{\boldsymbol{\beta}}_{j}^{(K-1)} = \arg\max_{\boldsymbol{\beta}_j}
        Q_j(\boldsymbol{\beta}_{j};\hat{\mathbf{\Gamma}}) 
        \quad \text{s.t. } \boldsymbol{\beta}_{j} \in \Theta_j(\mathcal C_j^{(K-1)}).\]
More specifically, we consider merging two adjacent subsets, with the intuition that $\beta_{jk}^*$s that share the same value should cluster together in $\hat{\boldsymbol\beta}_j$. We consider $K-1$ candidate partitions 
\begin{equation*}
\begin{aligned}
\mathcal C_{j1}^{(K-1)} &= \left\{\{\pi_j(1), \pi_j(2)\}, \{\pi_j(3)\}, ..., \{\pi_{j}(K)\}\right\},\\
\mathcal C_{j2}^{(K-1)} &= \left\{\{\pi_j(1)\}, \{\pi_j(2), \pi_j(3)\}, ..., \{\pi_{j}(K)\}\right\},\\
&...\\
\mathcal C_{j,K-1}^{(K-1)} &= \left\{\{\pi_j(1)\}, \{\pi_j(2)\}, ..., \{\pi_{j}(K-1), \pi_{j}(K)\}\right\},
\end{aligned}
\end{equation*}
 each of which contains $K-1$ subsets and is obtained by merging two adjacent subsets in $\mathcal C_j^{(K)}$. For each candidate partition, $\mathcal C_{jb}^{(K-1)}$, $b = 1, ..., K-1$, we compute a constrained solution
  \[\tilde{\boldsymbol{\beta}}_{jb}^{(K-1)} =
        \arg\max_{\boldsymbol{\beta}_j}
        Q_j(\boldsymbol{\beta}_{j};\hat{\mathbf{\Gamma}}) 
        \quad \text{s.t. } \boldsymbol{\beta}_j\in\Theta_j(\mathcal C_{jb}^{(K-1)}).\]
By construction, $\text{card}(\tilde{\boldsymbol{\beta}}_{jb}^{(K-1)}) = K-1$ for all $b$. The best $(K-1)$-level solution is given by the one that yields the largest pseudo-likelihood. That is,  
$\tilde{\boldsymbol{\beta}}_{j}^{(K-1)} = \tilde{\boldsymbol{\beta}}_{jb_{j}^{(K-1)}}^{(K-1)},$ 
with associated partition $\mathcal C_j^{(K-1)} = \mathcal C_{j, b_{j}^{(K-1)}}^{(K-1)}$, where 
\[b_{j}^{(K-1)} = \arg\max_{1\le b\le K-1} Q_j\!\left(\tilde{\boldsymbol{\beta}}_{jb}^{(K-1)};\hat{\mathbf{\Gamma}}\right).\]

Iteratively, suppose that the best $s$-level solution $\tilde{\boldsymbol{\beta}}_j^{(s)}$ has been obtained, with associated partition $\mathcal C_j^{(s)} = \{C_{j1}^{(s)},\ldots,C_{js}^{(s)}\}$, for $s  = 2, 3, ..., K$. 
The derivation of the best $(s-1)$-level solution $\tilde{\boldsymbol{\beta}}_j^{(s-1)}$ is similar to that of $\tilde{\boldsymbol{\beta}}_j^{(K-1)}$. That is, the candidate partitions for the $(s-1)$-level solution are given by 
\begin{equation*}
\begin{aligned}
\mathcal C_{j1}^{(s-1)} &= \left\{C_{j1}^{(s)}\cup C_{j2}^{(s)},C_{j3}^{(s)},\ldots,C_{js}^{(s)}\right\},\\
\mathcal C_{j2}^{(s-1)} &= \left\{C_{j1}^{(s)}, C_{j2}^{(s)}\cup C_{j3}^{(s)},\ldots,C_{js}^{(s)}\right\},\\
&...\\
\mathcal C_{j,s-1}^{(s-1)} &= \left\{C_{j1}^{(s)}, {C_{j2}^{(s)}}, \cdots,  C_{j,s-1}^{(s)}\cup C_{js}^{(s)}\right\}.
\end{aligned}
\end{equation*}
For each candidate partition, we compute a constrained solution
  \[\tilde{\boldsymbol{\beta}}_{jb}^{(s-1)} =
        \arg\max_{\boldsymbol{\beta}_j}
        Q_j(\boldsymbol{\beta}_{j};\hat{\mathbf{\Gamma}}) 
        \quad \text{s.t. } \boldsymbol{\beta}_j\in\Theta_j(\mathcal C_{jb}^{(s-1)}), ~b = 1, ..., s-1.\]
Then, 
$\tilde{\boldsymbol{\beta}}_{j}^{(s-1)} = \tilde{\boldsymbol{\beta}}_{jb_{j}^{(s-1)}}^{(s-1)},$ 
with associated partition $\mathcal C_j^{(s-1)} = \mathcal C_{j b_{j}^{(s-1)}}^{(s-1)}$,
where  
\[b_{j}^{(s-1)} = \arg\max_{1\le b\le s-1} Q_j\!\left(\tilde{\boldsymbol{\beta}}_{jb}^{(s-1)};\hat{\mathbf{\Gamma}}\right).\]

By running the above calculation for $s = K, K-1, ..., 2$, we obtain $\tilde{\boldsymbol{\beta}}_{j}^{(1)}$, ..., $\tilde{\boldsymbol{\beta}}_{j}^{(K)}$ as the candidate 1- to $K$-level models. 
This process is summarised in Algorithm~\ref{alg:stepwise}. 
Compared with an exhaustive search over all possible sparse models, the stepwise algorithm above is significantly more computationally efficient. That is, Algorithm~\ref{alg:stepwise} compares only $K(K-1)/2$ models, whereas the exhaustive search compares a number of models that is superexponential in $K$. When $K=10$, the stepwise method reduces the number of models to compare from 115,975 in the exhaustive search to 45.

 Although the algorithm does not exhaustively search over all possible sparse models, the theoretical results in Section~\ref{subsec:theory} show that, as the sample size $N$ goes to infinity, the true sparse structure for item $j$ is captured by $\tilde{\boldsymbol{\beta}}_{j}^{(m_j^*)}$ with probability tending to 1. Consequently, together with the EBIC introduced below, consistent model selection can be achieved.

\begin{algorithm}[h]
\caption{Stepwise search for item-level sparsity}\label{alg:stepwise}
\begin{algorithmic}[1]
\State \textbf{Input:} Item $j$; number of latent classes $K$; unrestricted item-response estimate $\hat{\boldsymbol{\beta}}_j$; pseudo-likelihood function $Q_j(~\cdot~ ; \hat{\mathbf{\Gamma}})$. 
\State Let $\pi_j$ be a permutation of $\{1,\ldots,K\}$ such that $\hat\beta_{j,\pi_j(1)} <  \cdots < \hat\beta_{j,\pi_j(K)}$.
\State Set $\tilde{\boldsymbol{\beta}}_{j}^{(K)}\gets \hat{\boldsymbol{\beta}}_j$.
\State Initialise the ordered partition,  $\mathcal C_j^{(K)}\gets \bigl\{\{\pi_j(1)\},\ldots,\{\pi_j(K)\}\bigr\}$.
\For{$s=K,K-1,\ldots,2$}
    \State Write the current ordered partition as $\mathcal C_j^{(s)}=\{C_{j1}^{(s)},\ldots,C_{js}^{(s)}\}$.
    \For{$b=1,\ldots,s-1$}
        \State Form $\mathcal C_{jb}^{(s-1)}$ by merging the adjacent blocks $C_{jb}^{(s)}$ and $C_{j,b+1}^{(s)}$ in $\mathcal C_j^{(s)}$.
        \State Compute
        \[\tilde{\boldsymbol{\beta}}_{jb}^{(s-1)} =
        \arg\max_{\boldsymbol{\beta}_j}
        Q_j(\boldsymbol{\beta}_{j};\hat{\mathbf{\Gamma}}) 
        \quad \text{s.t. } \boldsymbol{\beta}_j\in\Theta_j(\mathcal C_{jb}^{(s-1)}).\]
    \EndFor
    \State Select $b_{j}^{(s-1)} = \arg\max_{1\le b\le s-1} Q_j\!\left(\tilde{\boldsymbol{\beta}}_{jb}^{(s-1)};\hat{\mathbf{\Gamma}}\right)$.
    \State Set $\tilde{\boldsymbol{\beta}}_{j}^{(s-1)}\gets \tilde{\boldsymbol{\beta}}_{jb_{j}^{(s-1)}}^{(s-1)}$.
    \State Set $\mathcal C_j^{(s-1)}\gets \mathcal C_{jb_{j}^{(s-1)}}^{(s-1)}$ and keep the inherited block order.
\EndFor
\State \textbf{Output:} Candidate solutions $\{\tilde{\boldsymbol{\beta}}_{j}^{(m)}:m=1,\ldots,K\}$.
\end{algorithmic}
\end{algorithm} 

\begin{examplecont}
{To further illustrate Algorithm~\ref{alg:stepwise}, we use a simulated dataset generated from the toy example in Table~\ref{tab:example}, with sample size $N=1000$. We consider Item~1, for which the true item response probability vector $\boldsymbol{\beta}_1^*$ is $(0.10,0.10,0.10,0.90)$. The initial estimate gives $\tilde {\boldsymbol{\beta}}_1^{(4)} = \hat{\boldsymbol{\beta}}_1 = (0.141, 0.103, 0.083, 0.883)$, where the estimates of the first three parameters are close to each other but not exactly the same. Sorting the entries of $\hat{\boldsymbol{\beta}}_1$ in increasing order gives $\pi_1(1)=3$, $\pi_1(2)=2$, $\pi_1(3)=1$, and $\pi_1(4)=4$, as well as the ordered partition
\(
\mathcal C_1^{(4)}=\bigl\{\{3\},\{2\},\{1\},\{4\}\bigr\}.
\)

To find $\tilde {\boldsymbol{\beta}}_1^{(3)},$ there are three candidate partitions obtained by merging adjacent blocks,
\begin{equation*}
\mathcal C_{11}^{(3)}=\bigl\{\{2,3\},\{1\},\{4\}\bigr\},\quad
\mathcal C_{12}^{(3)}=\bigl\{\{3\},\{1,2\},\{4\}\bigr\},\quad
\mathcal C_{13}^{(3)}=\bigl\{\{3\},\{2\},\{1,4\}\bigr\}.
\end{equation*}
These partitions yield the constrained estimates
\(
\tilde {\boldsymbol{\beta}}_{11}^{(3)}
=(0.141,\,0.094,\,0.094,\,0.883),
\)
\(
\tilde {\boldsymbol{\beta}}_{12}^{(3)}
=(0.124,\,0.124,\,0.083,\,0.883),
\)
and
\(
\tilde {\boldsymbol{\beta}}_{13}^{(3)}
=(0.490,\,0.103,\,0.083,\,0.490),
\)
with pseudo-likelihood values $-353.713$, $-354.332$, and $-532.043$, respectively. Since $\tilde {\boldsymbol{\beta}}_{11}^{(3)}$ gives the largest pseudo-likelihood, we obtain $\tilde {\boldsymbol{\beta}}_1^{(3)} = \tilde {\boldsymbol{\beta}}_{11}^{(3)}=(0.141,\,0.094,\,0.094,\,0.883)$,
with associated partition
\(
\mathcal C_1^{(3)}=\mathcal C_{11}^{(3)}=\bigl\{\{2,3\},\{1\},\{4\}\bigr\}.
\)

By further executing Algorithm~\ref{alg:stepwise}, we obtain 
\(
\tilde {\boldsymbol{\beta}}_1^{(2)}
=(0.114,\,0.114,\,0.114,\,0.883)
\)
and
\(
\tilde {\boldsymbol{\beta}}_1^{(1)}
=(0.323,\,0.323,\,0.323,\,0.323).
\)
We note that when the number of levels is correctly set at 2, the estimate $\tilde {\boldsymbol{\beta}}_1^{(2)}$ recovers the same sparsity pattern as the true parameter in Table~\ref{tab:example}, which is $\mathcal C_{1}^{(2)} = \{\{1,2,3\},\{4\}\}$.}
\end{examplecont}

\medskip
\noindent
{\bf Extended BIC.} To determine the optimal sparsity level for each item, we construct an Extended BIC (EBIC) as 
\begin{equation}\label{eq:BIC}
    \text{EBIC}_{j}(m_j) = -2Q_j(\boldsymbol{\tilde{\beta}}_{j}^{(m_j)} ; \hat{\mathbf{\Gamma}}) + m_j\log(N) + 2m_j \log(\rho). 
\end{equation}
This EBIC is constructed following the idea proposed in the EBIC for high-dimensional regression models \citep{chen2008extended,chen2012extended}, where $\rho$ is a constant that characterises the user's prior belief on model sparsity.  More specifically, the first two terms in \eqref{eq:BIC}
mimic the standard BIC \citep{schwarz1978estimating}, but replace a real likelihood function with the pseudo-likelihood. These terms arise from the Laplace approximation to the marginal likelihood in a Bayesian formulation in which a prior distribution is imposed on all possible models and their model parameters. 
The last term arises from the prior distribution over models with different sparsity levels, with $\rho \geq 1$ being the ratio of the prior probability of a randomly chosen $m$-level model to that of a randomly chosen $(m+1)$-level model. 
We set $\rho \geq 1$ to ensure that sparse models are preferred. When $\rho$ is set to be 1, the third term becomes zero, so that the EBIC becomes a pseudo-BIC. The role of $\rho$ will be clarified in Section 3.3.

Based on the EBIC, we select the number of distinct levels for item $j$ as
\begin{equation}\label{eq:select}
  \tilde m_j = \arg\min_{m_j} \text{EBIC}_{j}(m_j).  
\end{equation} 
Given $\tilde m_j$, we infer the sparse item-response probability pattern based on the refined estimate $\boldsymbol{\tilde{\beta}}_{j}^{(\tilde{m}_{j})}$.

Next, we comment on the EBIC. As we set $\rho$ to be a constant, the third term in \eqref{eq:BIC} is dominated by the second term $m_j\log(N)$ as the sample size $N$ goes to infinity. Consequently, it will not affect the asymptotic property of the EBIC. However, this term is useful when the number of classes is relatively large compared with the sample size. In that case, the model space is large, and the effective sample size is relatively small. As a result, the standard BIC tends to over-select, as in high-dimensional regression settings \citep{chen2008extended,chen2012extended}. Our simulation results in Section~\ref{sec:simu} show that this additional penalty term substantially improves model selection accuracy. In particular, $\rho=20$ provides a practical finite-sample compromise: it reduces over-selection relative to weaker penalties while avoiding the pronounced under-selection induced by larger penalties. Moreover, as shown in Section~\ref{subsec:theory}, the EBIC consistently learns the true number of levels $m_j^*$ for each item $j$, that is, as the sample size $N$ goes to infinity, with probability tending to 1, $\tilde  m_j = m_j^*$. 
Combined with the model selection consistency of $\tilde{\boldsymbol{\beta}}_{j}^{(m_j^*)}$ from the stepwise search algorithm, the refined estimator $\boldsymbol{\tilde{\beta}}_{j}^{(\tilde{m}_{j})}$
consistently recovers the sparse pattern of $\boldsymbol{\beta}^*_j$.

\begin{examplecont}
{We illustrate the use of EBIC by continuing the analysis of Example~\ref{exmp:1}. From the stepwise search, we obtained the four candidate models
with pseudo-likelihood values
\[
Q_1\!(\tilde{\boldsymbol{\beta}}_1^{(4)}\!;\hat{\mathbf\Gamma})=-353.47,\ 
Q_1\!(\tilde{\boldsymbol{\beta}}_1^{(3)}\!;\hat{\mathbf\Gamma})=-353.71,\ 
Q_1\!(\tilde{\boldsymbol{\beta}}_1^{(2)}\!;\hat{\mathbf\Gamma})=-355.60,\ 
Q_1\!(\tilde{\boldsymbol{\beta}}_1^{(1)}\!;\hat{\mathbf\Gamma})=-629.11.
\]
With $N=1000$ and $\rho=20$, the EBIC values for the four candidate models are
\[
\mathrm{EBIC}_1(4)=758.54,\quad
\mathrm{EBIC}_1(3)=746.12,\quad
\mathrm{EBIC}_1(2)=736.99,\quad
\mathrm{EBIC}_1(1)=1271.12.
\]
Hence,
\[
\tilde m_1=\arg\min_{m\in\{1,2,3,4\}} \mathrm{EBIC}_1(m)=2,
\]
which correctly recovers the true number of distinct levels for Item 1. And the corresponding estimate
\(
\tilde{\boldsymbol{\beta}}_1^{(2)}=(0.114,\,0.114,\,0.114,\,0.883)
\)
recovers the same sparsity pattern as the true parameter vector in Table~\ref{tab:example}.}
\end{examplecont}

\medskip
\noindent
{\bf Final estimator.}
We obtain the final estimate of the latent class model via constrained marginal maximum likelihood estimation, using the item-level clustering pattern determined by EBIC. 
That is, 
\begin{equation}
    (\bar{\boldsymbol\nu}, \bar{\mathbf B}) = \arg\max_{(\boldsymbol{\nu}, \mathbf B)} ~l(\boldsymbol{\nu}, \mathbf B), s.t. \beta_{jg} = \beta_{jh} \mbox{~if~} \tilde \beta_{jg}^{(\tilde{m}_{j})} = \tilde \beta_{jh}^{(\tilde{m}_{j})}, j=1, ..., J, \mbox{~and~} h,g = 1, ..., K.
\end{equation}
The standard error for each $\bar \beta_{jk}$ and $\bar \nu_k$ is also obtained based on the Fisher information matrix of this marginal likelihood.  
We summarise the full proposed procedure in Algorithm~\ref{alg:summary}. 

{\begin{algorithm}
\caption{Proposed refinement procedure}\label{alg:summary}
\begin{algorithmic}[1]
\State \textbf{Input:} Binary response matrix $\mathbf Y$; number of latent classes $K$; initial LCA estimate $(\hat{\boldsymbol{\nu}},\hat{\mathbf B})$; posterior probability matrix $\hat{\mathbf \Gamma}$; EBIC tuning parameter $\rho$.
\For{$j=1,\ldots,J$}
    \State Construct the item-specific pseudo-likelihood $Q_j(\boldsymbol{\beta}_j;\hat{\mathbf \Gamma})$ as in \eqref{eq:pseudo-lik}.
    \State Apply Algorithm~\ref{alg:stepwise} to obtain candidate estimates
    $\{\tilde{\boldsymbol{\beta}}_j^{(m)}:m=1,\ldots,K\}$.
    \For{$m=1,\ldots,K$}
        \State Compute
        \[
        \operatorname{EBIC}_j(m)
        =
        -2Q_j(\tilde{\boldsymbol{\beta}}_j^{(m)};\hat{\mathbf \Gamma})
        +m\log N+2m\log \rho .
        \]
    \EndFor
    \State Select
    \[
    \tilde m_j = \arg\min_{1\le m\le K}\operatorname{EBIC}_j(m),
    \]
    and let $\tilde{\mathcal C}_j$ be the partition induced by
    $\tilde{\boldsymbol{\beta}}_j^{(\tilde m_j)}$.
\EndFor
\State Estimate $(\bar{\boldsymbol{\nu}},\bar{\mathbf B})$ by maximising the marginal likelihood subject to $\boldsymbol{\beta}_j\in\Theta_j(\tilde{\mathcal C}_j)$ for all $j=1,\ldots,J$.
\State \textbf{Output:} Selected item-level sparsity pattern $\{\tilde m_j,\tilde{\mathcal C}_j:j=1,\ldots,J\}$; final sparse LCA estimate $(\bar{\boldsymbol{\nu}},\bar{\mathbf B})$.
\end{algorithmic}
\end{algorithm}}

\subsection{Theoretical Results} \label{subsec:theory}

In what follows, we show that the sparse item-response probability pattern
implied by $\boldsymbol{\tilde{\beta}}_{j}^{(\tilde{m}_{j})}$ for each item $j$ is statistically consistent, for all $j = 1, ..., J$. This result further implies that 
the statistical inference based on the final estimate $(\bar{\mathbf B}, \bar{\boldsymbol\nu})$ is asymptotically valid. We make the following assumptions. 

\begin{assumption}\label{assm:MLE-consistency}
The MLE in \eqref{eq:MLE} satisfies 
\begin{equation}
\Vert \hat{\boldsymbol{\nu}} - {\boldsymbol{\nu}}^*\Vert = O_p(1/\sqrt{N}) \mbox{~and~} \Vert \hat{\mathbf B} - {\mathbf B}^* \Vert_F = O_p(1/\sqrt{N}),
\end{equation}
where ${\boldsymbol{\nu}}^*$ and ${\mathbf B}^*$ are the corresponding true model parameters.
\end{assumption}
 
\begin{assumption}\label{assm:inner}

The true parameters lie in the interior of the parameter space; that is, $0 < \beta_{jk}^* < 1$ and $\nu_k^* > 0$ for all $j=1,\dots,J$ and $k=1,\dots,K$.
\end{assumption}

Since $J$ and $K$ are fixed, Assumption~2 implies that there exists a constant $\varepsilon>0$ such that $\varepsilon\le \beta_{jk}^*\le 1-\varepsilon$ and $\nu_k^*\ge \varepsilon$ for all $j=1,\dots,J$ and $k=1,\dots,K$.

\begin{assumption}\label{assm:rho}
The parameter $\rho$ in the EBIC is a constant that does not depend on the sample size $N$. 
\end{assumption}

\begin{theorem}\label{thm:main}
Suppose that Assumptions 1--3 hold. Then, for each item $j = 1, ..., J$, the probability $P(\tilde m_j = m_j^*)$ converges to 1, as the sample size $N$ goes to infinity. Moreover, for each item $j = 1, ..., J$, the refined estimator $\boldsymbol{\tilde{\beta}}_{j}^{(\tilde{m}_{j})}$ satisfies: 
\begin{enumerate}
    \item  
$P(\tilde{\beta}_{jg} = \tilde{\beta}_{jh})$ converges to 1, for all $g$ and $h \in \{1, ..., K\}$ such that $\beta_{jg}^* = \beta_{jh}^*$ , and 
\item $P(\tilde{\beta}_{jg} \neq \tilde{\beta}_{jh})$ converges to 1, for all $g$ and $h \in \{1, ..., K\}$ such that $\beta_{jg}^* \neq \beta_{jh}^*$,
\end{enumerate}
as the sample size $N$ goes to infinity. 
\end{theorem}
The proof of the theorem is given in Section~\ref{S-app:proof-of-theorem} of the Supplementary Materials. Theorem~\ref{thm:main} gives two results about the statistical consistency of the proposed method. The first result, $P(\tilde m_j=m_j^*)$ converges to 1, states that the EBIC selection step consistently identifies the true number of distinct response-probability levels for each item. The second result, which concerns $P(\tilde{\beta}_{jg} = \tilde{\beta}_{jh})$ and $P(\tilde{\beta}_{jg} \neq  \tilde{\beta}_{jh})$,
further establishes the consistency in recovering the sparse pattern for each item: latent classes with the same true item-response probability are asymptotically grouped together, whereas latent classes with different true item-response probabilities are asymptotically kept in separate groups. Thus, the refinement procedure consistently recovers both the item-specific model complexity and the equality pattern among the $K$ latent classes.

\section{Simulation Study}\label{sec:simu}

\subsection{Simulation Settings}

We consider two simulation settings. Setting I has $K=4$ latent classes and $J=32$ binary items, and Setting II has $K=8$ latent classes and $J=64$ binary items. For each setting, we generate one true class proportion vector $\boldsymbol{\nu}$ and one true item-response probability matrix $\mathbf B$, and then keep these true parameters fixed across all sample sizes and replications. The true class proportions are $\boldsymbol{\nu}=(0.241,0.259,0.202,0.298)$ in Setting I and $\boldsymbol{\nu}=(0.106,0.137,0.130,0.101,0.124,0.117,0.150,0.136)$ in Setting II. For each setting, we consider sample sizes $N\in\{500,750,1000,1500,2000\}$ and run 100 independent replications for each sample size.
The two settings also differ in the level configurations of the true item-response probabilities. \rv{In Setting I, all 32 items have two distinct response-probability levels, where a level denotes a distinct value among the \(K\) class-specific response probabilities for a given item rather than an observed response category. In Setting II, the 64 items consist of 48 two-level items and 16 three-level items, yielding a richer and more complex item-level structure.} The full true class proportions and item-response probabilities are reported in Supplementary Section~\ref{S-app:true-values}.

In the refinement stage, we consider multiple candidate values of the EBIC tuning parameter $\rho$, namely 1, 5, 10, 20, 40, 80, 160, and 320. Recall that the additional EBIC penalty is $2m_j\log\rho$, so this term is zero when $\rho=1$ and increases with $\rho$, placing a stronger penalty on denser models.

In the main simulation analysis, the unrestricted first-stage model for $(\hat{\mathbf B},\hat{\boldsymbol{\nu}})$ is fitted using the true value of $K$, so that the reported results focus on the proposed item-level refinement.\footnote{Additional simulations have been conducted to assess the selection of the number of classes $K$ by BIC. The results, which are given in Section~\ref{S-app:bic-k-selection} of the Supplementary Materials, show that $K$ can be accurately selected under the current simulation settings.}\rv{ Furthermore, we also conducted additional simulations to compare the proposed procedure with the regularised method of \citet{chen2017regularized}; the results are reported in Section~\ref{S-sec:chen2017-comparison} of the Supplementary Materials.}

\subsection{Evaluation Criteria}

{Multiple criteria are used to evaluate the performance of the proposed method. First, we evaluate the selection of the item-specific numbers of probability levels. For each simulated dataset, we record the numbers of items that are under-selected, correctly selected, and over-selected relative to the true values $\{m_j^*\}_{j=1}^J$. We also record the total number of incorrectly selected items per replication, where an item is counted as incorrect whenever $\tilde m_j \neq m_j^*$.

Second, we assess recovery of the sparse item-response probability pattern itself. This is done by comparing the partition of the latent classes induced by $\boldsymbol{\beta}_j^*$, denoted $\mathcal C_j^*$, with that induced by $\tilde{\boldsymbol{\beta}}_j$, denoted $\tilde{\mathcal{C}}_j$, using the adjusted Rand index (ARI; \citealp{Hubert1985}). For each item $j$, we compute
\[
\mathrm{ARI}(\mathcal C_j^*,\tilde{\mathcal C}_j)
=
\frac{\sum_{g,h}\binom{n_{gh}}{2}-\frac{\sum_g\binom{n_{g\cdot}}{2}\sum_h\binom{n_{\cdot h}}{2}}{\binom{K}{2}}}
{\frac12\left[\sum_g\binom{n_{g\cdot}}{2}+\sum_h\binom{n_{\cdot h}}{2}\right]-\frac{\sum_g\binom{n_{g\cdot}}{2}\sum_h\binom{n_{\cdot h}}{2}}{\binom{K}{2}}}.
\]
Here, $n_{gh}$ denotes the number of latent classes that fall into group $g$ under $\mathcal C_j^\star$ and into group $h$ under $\tilde{\mathcal C}_j$, and $n_{g\cdot}=\sum_h n_{gh}$ and $n_{\cdot h}=\sum_g n_{gh}$ are the corresponding marginal counts. For each simulated dataset and each value of $\rho$, we compute the average item-level ARI, $(1/J)\sum_{j=1}^J \mathrm{ARI}(\mathcal C_j^\star,\tilde{\mathcal C}_j)$.

Finally, we assess parameter estimation accuracy for both the item-response probabilities and the class proportions. Specifically, for each simulated dataset and each value of $\rho$, we compute the mean squared error of the final refined estimator,
\[
\mathrm{MSE}(\bar{\mathbf B})
=
\frac{1}{JK}\min_{\pi}\sum_{j=1}^J\sum_{k=1}^K\left(\bar{\beta}_{j,\pi(k)}-\beta_{jk}^*\right)^2,\]
where the minimum is taken over all permutations $\pi$ on $\{1, ..., K\}$ to account for label indeterminacy. Let $\hat{\pi}$ denote the permutation that attains this minimum. We then compute the corresponding error for the class proportions using the same alignment,
\[
\mathrm{MSE}(\bar{\boldsymbol{\nu}})
=
\frac{1}{K}\sum_{k=1}^K\left(\bar{\nu}_{\hat{\pi}(k)}-\nu_k^*\right)^2.
\]
To evaluate the effect of the refinement step, we compute the same quantities for the unrestricted first-stage estimator $(\hat{\mathbf B},\hat{\boldsymbol{\nu}})$, and compare them with those of the final refined estimator $(\bar{\mathbf B},\bar{\boldsymbol{\nu}})$.}

\subsection{Simulation Results}\label{subsec:sim-results}
{We first discuss the choice of $\rho$, which controls a clear trade-off between over-selection and under-selection in the refinement step. As $\rho$ increases, denser models are penalised more heavily, so over-selection tends to decrease. At the same time, a larger $\rho$ makes the procedure more prone to under-selection.
Table~\ref{tab:sim-rho-tradeoff} shows the average numbers of under-selected and over-selected items under different values of $\rho$. In Setting I, this trade-off is mild. Because this setting is relatively simple, the overall performance is strong across all sample sizes. Under-selection is essentially absent across all sample sizes and all candidate values of $\rho$, while over-selection decreases steadily as $\rho$ increases and also becomes smaller as $N$ increases. In contrast, the trade-off is much more pronounced in Setting II. For each fixed $\rho$, both types of selection error generally decrease as $N$ increases. For each fixed sample size, increasing $\rho$ sharply reduces over-selection, but this improvement is accompanied by more under-selection, especially when $N$ is small. For example, at $N=500$, over-selection drops from 8.94 items at $\rho=1$ to 0.40 at $\rho=20$, while under-selection increases from 0.27 to 2.33.

\begin{table}[ht]
\centering
\caption{Item-level selection errors under different EBIC tuning values $\rho$. For each setting and sample size, the table reports the average numbers of under-selected and over-selected items per replication. The row for $\rho=20$ is highlighted because this value is used in the remaining simulation summaries.}
\label{tab:sim-rho-tradeoff}
\begingroup
\small
\setlength{\tabcolsep}{5pt}
\renewcommand{\arraystretch}{1.12}
\begin{tabularx}{\textwidth}{r *{5}{Y} *{5}{Y}}
\toprule
\multicolumn{11}{l}{\textbf{Setting I}} \\
\multicolumn{1}{c}{$\rho$} & \multicolumn{5}{c}{Under-selected items} & \multicolumn{5}{c}{Over-selected items} \\
\cmidrule(lr){2-6}\cmidrule(lr){7-11}
& $500$ & $750$ & $1000$ & $1500$ & $2000$ & $500$ & $750$ & $1000$ & $1500$ & $2000$ \\
\midrule
1 & 0.00 & 0.00 & 0.00 & 0.00 & 0.00 & 1.45 & 1.32 & 1.06 & 0.76 & 0.69 \\
5 & 0.00 & 0.00 & 0.00 & 0.00 & 0.00 & 0.25 & 0.28 & 0.24 & 0.14 & 0.14 \\
10 & 0.00 & 0.00 & 0.00 & 0.00 & 0.00 & 0.11 & 0.19 & 0.13 & 0.08 & 0.10 \\
\textbf{20} & \textbf{0.00} & \textbf{0.00} & \textbf{0.00} & \textbf{0.00} & \textbf{0.00} & \textbf{0.07} & \textbf{0.11} & \textbf{0.08} & \textbf{0.06} & \textbf{0.06} \\
40 & 0.00 & 0.00 & 0.00 & 0.00 & 0.00 & 0.05 & 0.03 & 0.01 & 0.04 & 0.03 \\
80 & 0.00 & 0.00 & 0.00 & 0.00 & 0.00 & 0.04 & 0.01 & 0.00 & 0.03 & 0.02 \\
160 & 0.00 & 0.00 & 0.00 & 0.00 & 0.00 & 0.03 & 0.01 & 0.00 & 0.01 & 0.00 \\
320 & 0.00 & 0.00 & 0.00 & 0.00 & 0.00 & 0.03 & 0.00 & 0.00 & 0.01 & 0.00 \\
\midrule
\multicolumn{11}{l}{\textbf{Setting II}} \\
\multicolumn{1}{c}{$\rho$} & \multicolumn{5}{c}{Under-selected items} & \multicolumn{5}{c}{Over-selected items} \\
\cmidrule(lr){2-6}\cmidrule(lr){7-11}
& $500$ & $750$ & $1000$ & $1500$ & $2000$ & $500$ & $750$ & $1000$ & $1500$ & $2000$ \\
\midrule
1 & 0.27 & 0.02 & 0.00 & 0.00 & 0.00 & 8.94 & 7.13 & 6.64 & 5.47 & 4.25 \\
5 & 1.01 & 0.12 & 0.01 & 0.00 & 0.00 & 2.00 & 1.65 & 1.47 & 1.17 & 0.86 \\
10 & 1.45 & 0.17 & 0.01 & 0.00 & 0.00 & 1.09 & 0.71 & 0.68 & 0.69 & 0.45 \\
\textbf{20} & \textbf{2.33} & \textbf{0.32} & \textbf{0.01} & \textbf{0.00} & \textbf{0.00} & \textbf{0.40} & \textbf{0.41} & \textbf{0.36} & \textbf{0.38} & \textbf{0.21} \\
40 & 3.24 & 0.46 & 0.02 & 0.00 & 0.00 & 0.16 & 0.21 & 0.17 & 0.17 & 0.11 \\
80 & 4.04 & 0.64 & 0.06 & 0.00 & 0.00 & 0.09 & 0.07 & 0.08 & 0.11 & 0.08 \\
160 & 5.18 & 0.86 & 0.12 & 0.00 & 0.00 & 0.07 & 0.05 & 0.02 & 0.05 & 0.04 \\
320 & 6.24 & 1.28 & 0.14 & 0.00 & 0.00 & 0.03 & 0.03 & 0.01 & 0.04 & 0.01 \\
\bottomrule
\end{tabularx}
\endgroup
\end{table}}

For the remaining analyses, we use $\rho=20$.  This choice is motivated by
the fact that the real-data application in Section~\ref{sec:data} has sample and item sizes similar to those of the $N=750$ case under Setting II. For this case, 
 $\rho=20$ provides a reasonably balanced compromise between over-selection and under-selection.  In particular, at $N=750$ in Setting II, increasing $\rho$ from 1 to 20 reduces the average number of over-selected items from 7.13 to 0.41, while the average number of under-selected items is still moderate at 0.32; larger values of $\rho$ reduce over-selection only slightly further but induce more under-selection. This asymmetry is important because under-selection can introduce substantial bias in the refined estimates of $\bar{\mathbf B}$, whereas moderate over-selection is less harmful for estimation accuracy. We therefore use $\rho=20$ in the remaining simulation summaries and in the real-data analysis. Results for the other candidate values of $\rho$ are reported in Supplementary Section~\ref{S-app:rho-sensitivity}.

\begin{figure}[tbp]
\centering
\includegraphics[width=0.95\textwidth]{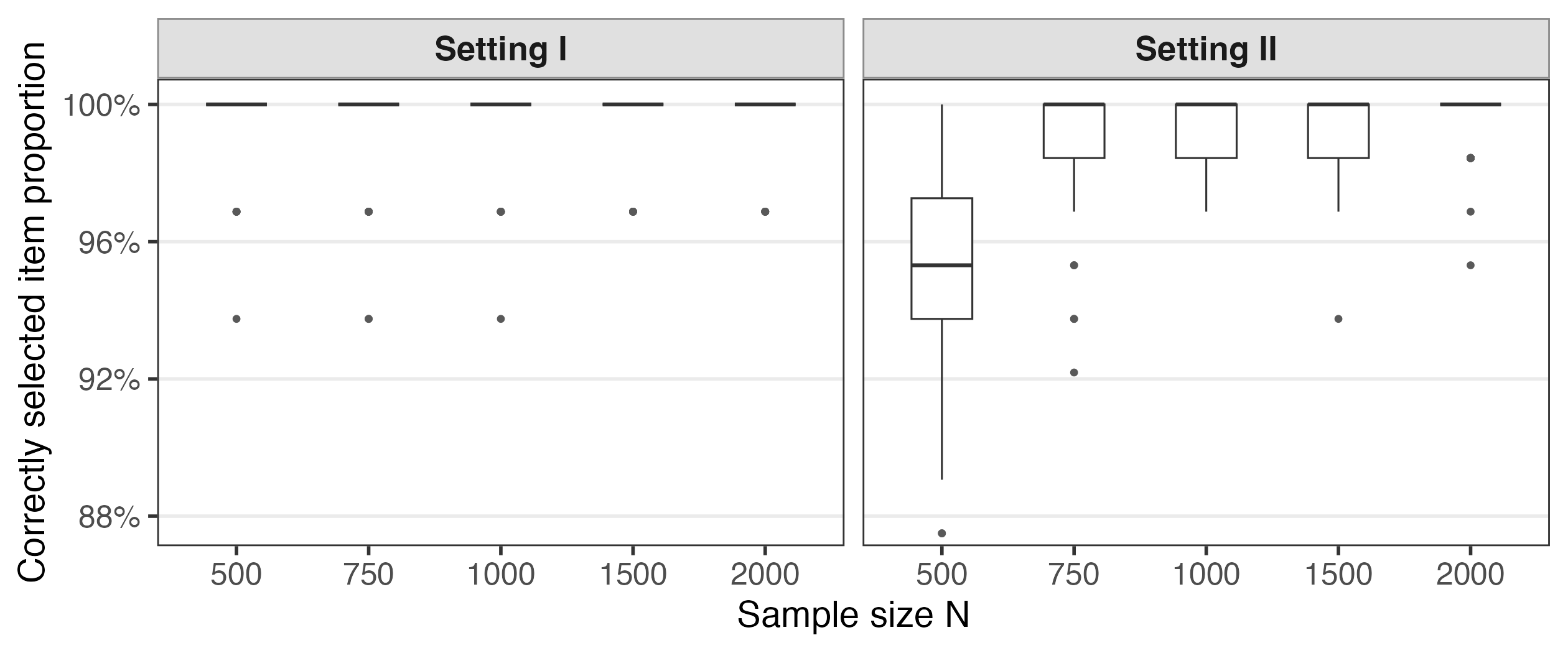}
\caption{Proportion of items with correctly selected $m_j$ in each replication at $\rho=20$. Columns correspond to Settings I and II.}
\label{fig:sim-recovery-errors}
\end{figure}

{Figure~\ref{fig:sim-recovery-errors} summarises, for each replication, the proportion of items for which the number of levels, $m_j$, is correctly selected. In Setting I, this proportion is concentrated very close to one across all sample sizes, indicating that EBIC almost always recovers the true item-level complexity. Even in the smallest samples, the worst replications incorrectly select at most two items out of 32, and the distribution becomes increasingly concentrated at perfect selection as $N$ increases. Setting II is more challenging, but the boxplots still shift steadily upward with $N$. At $N=500$, the median proportion of correctly selected items is already above 0.95, meaning that the typical replication incorrectly selects no more than three items out of 64. Once $N$ reaches 750 or above, the median is essentially at perfect selection, and by $N=2000$, most replications achieve perfect selection. Overall, these results show that the EBIC step accurately identifies the true number of item-level probability levels, and that its performance improves with sample size, in line with theoretical results.}

\begin{figure}[t]
\centering
\includegraphics[width=0.95\textwidth]{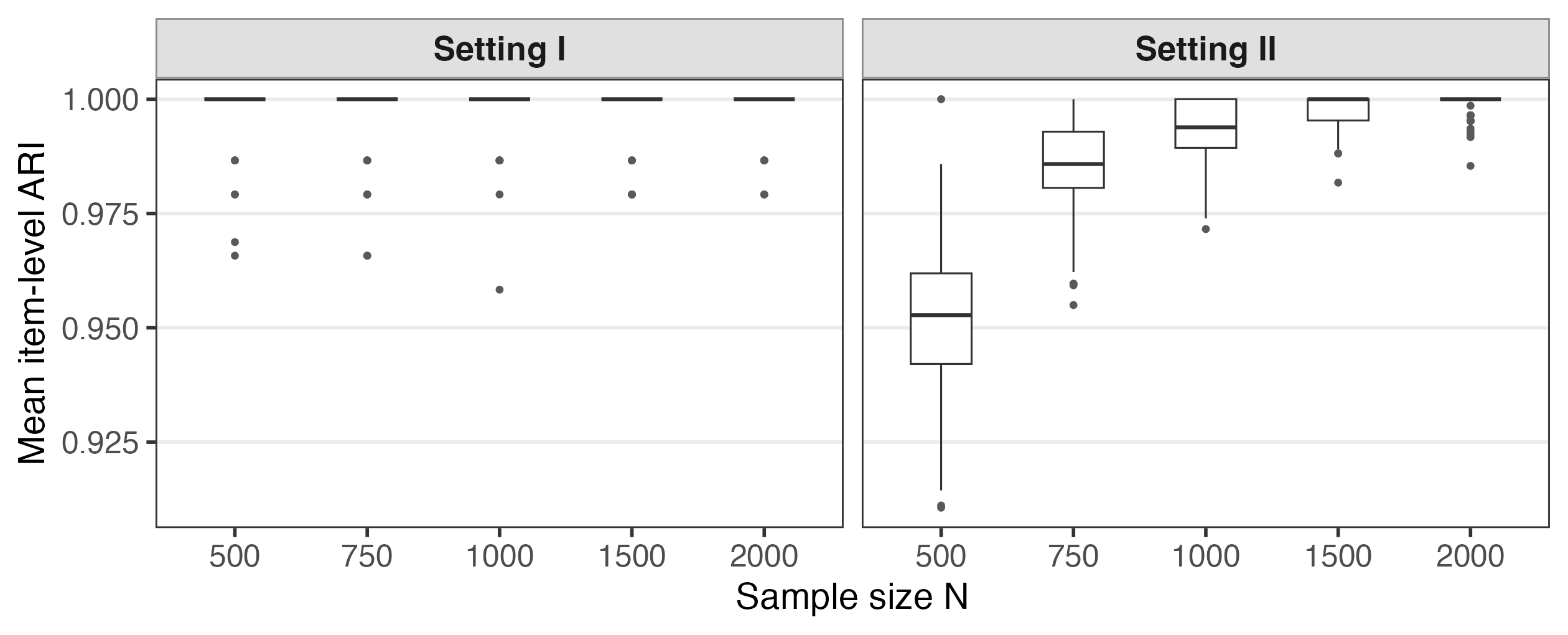}
\caption{Mean item-level ARI in each replication at $\rho=20$. Columns correspond to Settings I and II.}
\label{fig:sim-recovery-ari}
\end{figure}

{Figure~\ref{fig:sim-recovery-ari} provides a complementary summary based on the mean item-level ARI, which measures how well the estimated class partition matches the true sparse partition once $m_j$ has been selected. In Setting I, the ARI remains high, and the median is essentially at perfect partition recovery across all sample sizes; as $N$ increases, the remaining outliers become less frequent. In Setting II, recovery is weaker at $N=500$, with a mean item-level ARI of around 0.95, but it improves quickly with sample size and exceeds 0.98 once $N$ reaches 750. By $N=2000$, most replications recover the true sparse pattern almost perfectly, apart from a small number of outliers. Taken together, these results show that the selected item-level sparsity pattern converges rapidly to the truth, with most remaining recovery errors concentrated in the smallest-sample Setting II scenario.}

\begin{figure}[t]
\centering
\includegraphics[width=0.95\textwidth]{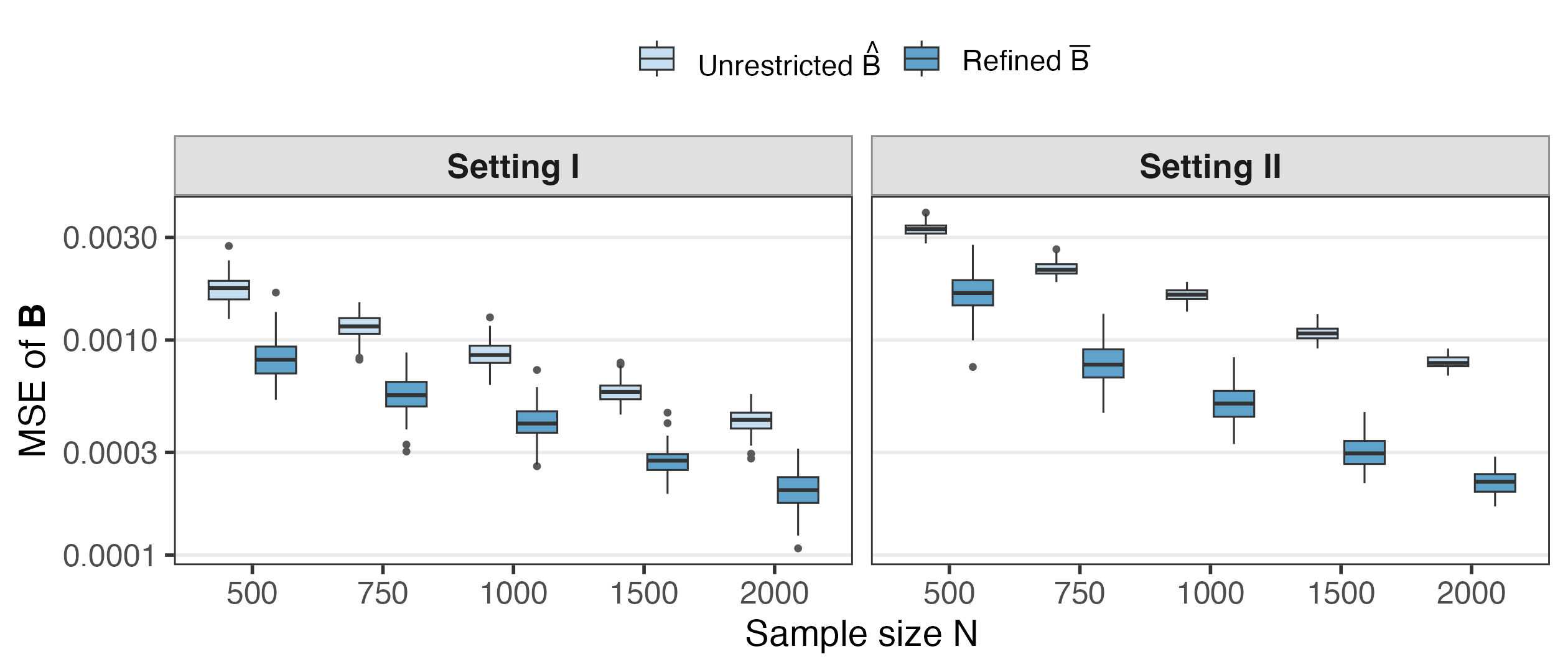}

\includegraphics[width=0.95\textwidth]{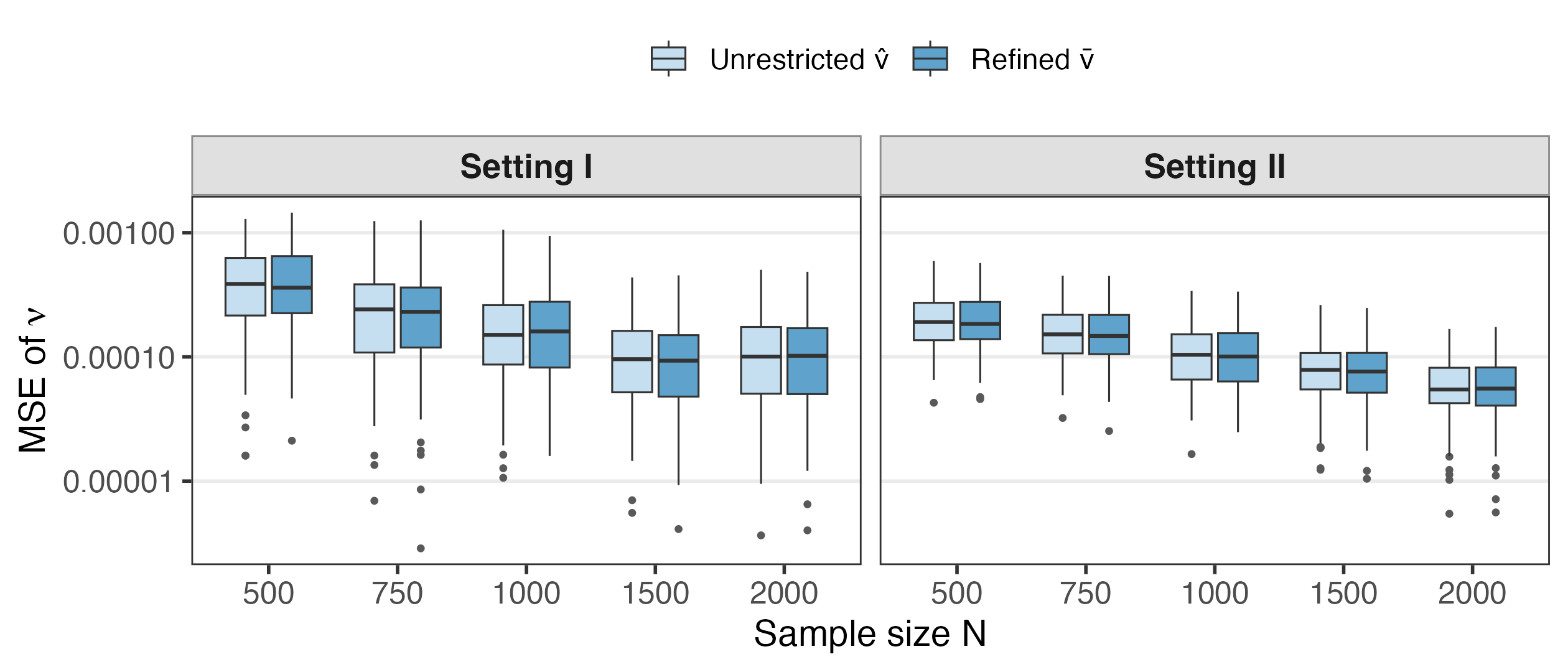}
\caption{Mean squared error of the item-response probability matrix $\mathbf B$ and class proportions $\boldsymbol{\nu}$ across replications at $\rho=20$. The top panel reports the MSE for $\mathbf B$, and the bottom panel reports the MSE for $\boldsymbol{\nu}$. Within each setting and sample size, the two boxplots compare the unrestricted first-stage estimate with the final refined estimate.}
\label{fig:sim-mse}
\end{figure}

{Figure~\ref{fig:sim-mse} compares the MSE of the unrestricted first-stage estimate with that of the final refined estimate, separately for $\mathbf B$ and $\boldsymbol{\nu}$. For both parameters, the MSE decreases with increasing sample size, consistent with theoretical results. For $\mathbf B$, the refined estimator uniformly improves on the unrestricted estimator across both simulation settings and all sample sizes. In Setting I, the mean MSE decreases from $1.73\times 10^{-3}$ to $8.32\times 10^{-4}$ at $N=500$, and from $4.27\times 10^{-4}$ to $2.04\times 10^{-4}$ at $N=2000$. In Setting II, the corresponding decrease is from $3.27\times 10^{-3}$ to $1.69\times 10^{-3}$ at $N=500$, and from $7.90\times 10^{-4}$ to $2.20\times 10^{-4}$ at $N=2000$. These gains indicate that, once the item-level sparse structure is recovered with sufficient accuracy, imposing the selected equality constraints substantially improves estimation of the item-response probability matrix.}

For the class proportions $\boldsymbol{\nu}$, the unrestricted and refined estimates are much closer, as expected, because the refinement step is designed primarily to improve estimation of $\mathbf B$ rather than the class proportions. The MSE of $\boldsymbol{\nu}$ nevertheless decreases with sample size in both settings. In Setting I, the mean refined MSE decreases from $4.53\times 10^{-4}$ at $N=500$ to $1.18\times 10^{-4}$ at $N=2000$; in Setting II, it decreases from $2.14\times 10^{-4}$ to $6.23\times 10^{-5}$ over the same range. Thus, the main estimation gain from refinement occurs for $\mathbf B$, while estimation of $\boldsymbol{\nu}$ is largely preserved.

\section{An Application to Social Role Performance Data}\label{sec:data}

\subsection{Data}

{We applied the proposed refinement procedure to the Social Role Performance (SRPPER) item bank from the PROMIS Wave 1 calibration study \citep{Cella2010}. The analysis included respondents with complete responses to all SRPPER items, yielding a response matrix with $N=737$ respondents and $J=56$ items, listed in Supplementary Table~\ref{S-tab:real-item-blocks}.

The original SRPPER items in the PROMIS dataset were measured on a five-point ordinal scale with response categories Never, Rarely, Sometimes, Often, and Always. Because the items included both positively and negatively worded statements, the numeric scores from 1 to 5 were assigned to align all items in the same direction, with higher values indicating more severe limitations in social role performance. In the analysis, we dichotomised each item as $Y_{ij}=1$ for responses coded 4 or 5 and $Y_{ij}=0$ otherwise. Therefore, the item-response probabilities reported below can be interpreted as the probabilities of endorsing a more severe response under this binary coding.
}

\subsection{Analysis Workflow}

{The real-data analysis followed the strategy developed in Section~\ref{subsec:refinement}. We first fitted unrestricted latent class models with candidate numbers of classes ranging from 1 to 8 and selected $K=8$ latent classes based on BIC. We then applied the proposed item-level refinement procedure to this selected unrestricted solution. The refinement step used $\rho=20$, as suggested by the simulation results (see Section~\ref{subsec:sim-results} and Setting II in Table~\ref{tab:sim-rho-tradeoff}), to select item-specific equality groups among the eight latent classes. Finally, the sparse latent class model was re-estimated under these selected equality constraints, producing the refined item-response probability matrix $\bar{\mathbf B}$ and the estimated class proportions.}

\subsection{Results}

{The refinement substantially reduced the complexity of the item-response probability matrix. Among the 56 SRPPER items, the selected number of distinct probability levels was $\tilde m_j=3$ for 7 items, $\tilde m_j=4$ for 38 items, and $\tilde m_j=5$ for 11 items. Thus, instead of estimating eight unrelated class-specific probabilities for every item, the refined model represents each item using only three to five ordered probability levels. This reduction is directly tied to the goal of the proposed method: it preserves heterogeneity across items while making the class differences more interpretable. The unrestricted first-stage estimates, together with the corresponding comparison heatmap using the same item grouping, are reported in Supplementary Section~\ref{S-sec:real-initial-estimates}. Relative to the refined matrix, the unrestricted estimates are visibly noisier: many item-specific probabilities vary gradually across classes rather than collapsing into a small number of repeated levels, so the overall severity ordering remains visible but the domain-specific contrasts are much harder to identify directly from the raw estimates.

For presentation, the eight latent classes were labelled in ascending order of average item-response probability. The estimated class proportions after this relabelling are 0.142, 0.086, 0.075, 0.089, 0.116, 0.071, 0.117, and 0.302. The corresponding average item-response probabilities are 0.054, 0.234, 0.304, 0.509, 0.586, 0.751, 0.851, and 0.976, showing a clear ordering from the least severe to the most severe latent profiles.

Figure~\ref{fig:real-beta-heatmap} displays the final refined item-response probability matrix $\bar{\mathbf B}$. Compared with the unrestricted heatmap reported in Supplementary Section~\ref{S-sec:real-initial-estimates}, where the probabilities often vary gradually across classes and the relevant level patterns are difficult to see, the refined matrix makes the level structure visually transparent. Classes assigned to the same item-specific level have identical refined probabilities, and classes assigned to adjacent levels can be interpreted as having low, intermediate, or high probabilities of endorsing the severe response. Consequently, when the items are ordered by hierarchical clustering and the latent classes are ordered from the lowest to the highest class-average severe-response probability, the heatmap reveals clearer block structures across items.

\begin{figure}[ht]
\centering
\includegraphics[width=\textwidth]{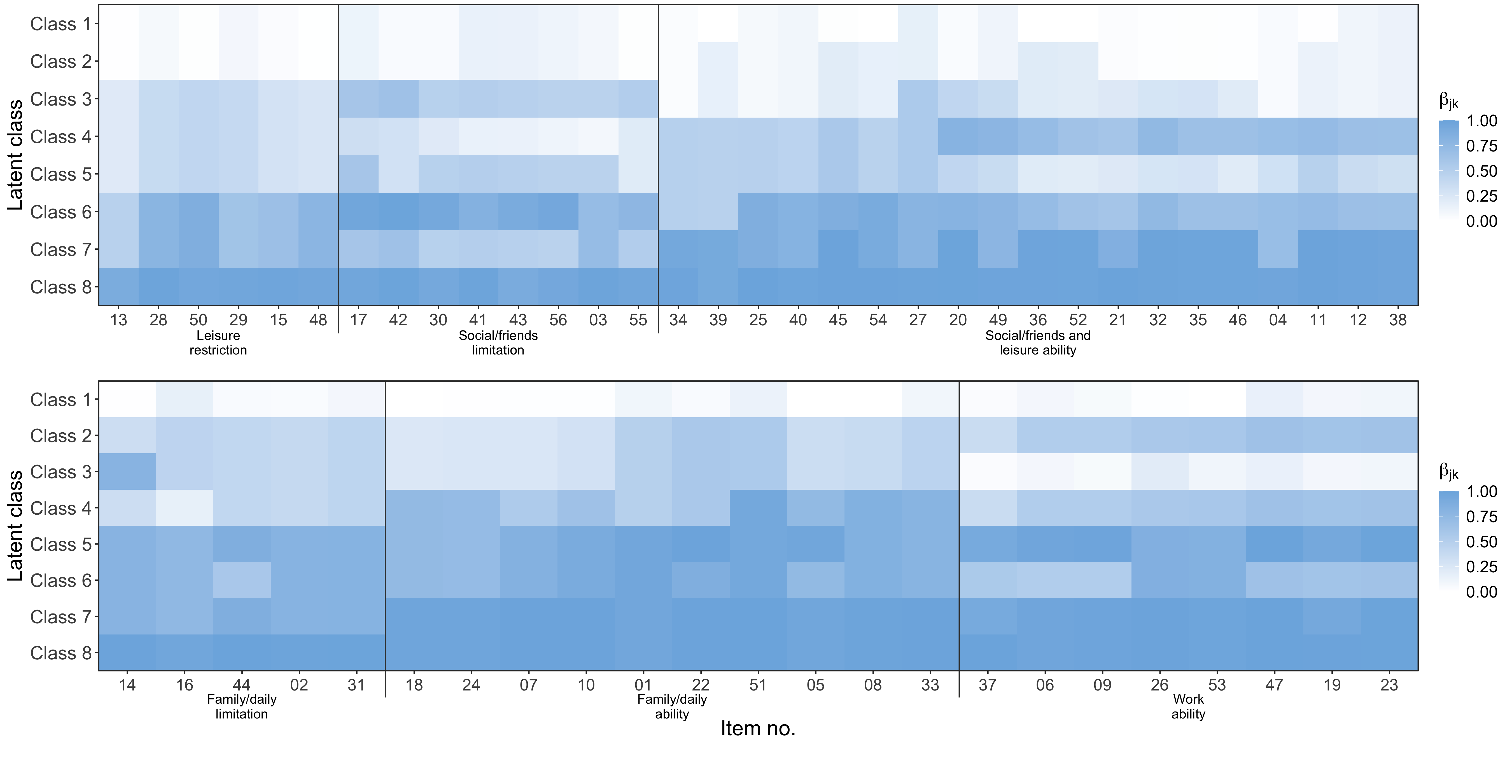}
\caption{Heatmap of the final refined item-response probability matrix for the PROMIS Social Role Performance data. For readability, the 56 items are displayed across two stacked panels. Within each block, items are ordered by hierarchical clustering of their refined probability profiles; the labels beneath the x-axis summarise the content of each block. Rows correspond to relabelled latent classes ordered by increasing average probability of endorsing the more severe binary response.}
\label{fig:real-beta-heatmap}
\end{figure}
}

{The hierarchical clustering of the items reveals six contiguous item blocks, highlighted in Figure~\ref{fig:real-beta-heatmap} and defined in Supplementary Table~\ref{S-tab:real-item-blocks}. They are (i) \emph{leisure restriction items} (6 items), (ii) \emph{social/friends limitation items} (8 items), (iii) \emph{social/friends and leisure ability items} (19 items), (iv) \emph{family and daily-role limitation items} (5 items), (v) \emph{family and daily-role ability items} (10 items), and (vi) \emph{work ability items} (8 items). This ordering is not driven by topical content alone. Instead, the clustering also separates item polarity: family and daily-role items split into limitation versus ability wording, while the broader social/leisure domain separates into a shared ability block and two negatively worded blocks corresponding primarily to leisure and to social/friends functioning.

Along the vertical axis, the relabelled Classes 1--8 form an overall severity ordering. Class~1 is characterised by the lowest or near-lowest item-specific probability level across essentially all blocks, indicating little endorsement of severe responses throughout leisure, social/friends, family/daily-role, and work domains. At the opposite end, Class~8 is uniformly high across all six blocks, with almost all items assigned to the highest or near-highest level, representing a globally severe profile.

The intermediate classes differ not only in overall severity but also in which domains become elevated first. In Class~2, the first three blocks, corresponding to leisure restriction, social/friends limitation, and social/friends and leisure ability, remain at low levels, whereas the family/daily-role and work blocks begin to rise. Class~3 shows a different pattern: the limitation blocks for social/friends and leisure become elevated, especially the social/friends limitation block, while the shared social/friends and leisure ability block remains relatively intact. In the family/daily-role domain, Class~3 is similar to Class~2, but its work block stays much closer to the low level. Class~4 is marked by a clearer increase in the family/daily-role ability block and in the social/friends and leisure ability block, suggesting that positive-functioning items in these domains are now differentiating the class more strongly than in the preceding profiles.

From Class~5 onward, the profiles move into a broadly moderate-to-high range. Class~5 is the first class showing a more general mid-to-high pattern across the heatmap, with the family/daily-role and work blocks rising earliest and most clearly. Class~6 is also moderate-to-high overall, but is distinguished by especially severe social/friends difficulties: the social/friends limitation block stands out more strongly, whereas the work block is not especially high relative to the other domains. Class~7 is higher still, with the three ability blocks---social/friends and leisure ability, family/daily-role ability, and work ability---already reaching the highest level for many items, while the two limitation-oriented domains remain mostly at intermediate-to-high levels. Thus, neighbouring classes are not separated by severity alone; they also differ in the domain-specific pattern of impairment and in whether the strongest differentiation appears first in limitation-oriented versus ability-oriented item blocks.

Overall, the refinement turns the dense $56\times 8$ probability matrix into a structured profile in which the latent classes differ not only in the severity and emphasis of social-role dysfunction, but also in their response tendency across positively versus negatively worded items. In particular, the heatmap suggests that some respondents may be relatively less likely to endorse the more severe wording on negatively framed limitation/restriction items while showing a different tendency on positively framed ability items. This means that the class structure reflects both substantive differences in social-role functioning and systematic differences in how respondents interact with item wording polarity across domains.
}

\section{Discussion}\label{sec:discussion}

In this paper, we introduce a post-estimation refinement procedure that bridges the gap between the exploratory power of LCA and its interpretability. By decoupling the initial estimation and the sparsity-inducing model selection, our approach sidesteps the computational bottlenecks with regularised estimation methods while maintaining the asymptotic guarantees necessary for reliable inference.
 
The proposed refinement procedure extends naturally to the LCA of polytomous data, where  
each item $j$ can have $C_j\geq 2$ response categories.
Under this setting, the item-response probability matrix expands into a three-way tensor, where each entry $\beta_{jkc}$ represents the probability of class $k$ providing response $c$ to item $j$. In this case, a refinement procedure would not only aim to collapse the $K$ multinomial distributions $\boldsymbol{\beta}_{jk} = (\beta_{jk1}, \dots, \beta_{jkC_j})^\top$, $k=1, ..., K$, into a set of $m_j < K$ unique distributions, but also collapse 
the multinomial probabilities $\beta_{jk1}, \dots, \beta_{jkC_j}$ into a smaller number of unique values for each class $k$. An item-specific pseudo-likelihood can be easily defined for this setting by replacing the weighted Bernoulli log-likelihood in \eqref{eq:pseudo-lik} with a weighted multinomial log-likelihood, based on which an EBIC can be defined accordingly. A new stepwise search algorithm is needed to accommodate the dual collapsing tasks -- one over the latent classes and the other over the response categories. One way to achieve it is to perform the search in two stages. In the first stage, we use a stepwise search algorithm, as in Algorithm 1, to collapse the multinomial distributions over the $K$ latent classes. In the second stage, for each collapsed class from the first stage, we further collapse the item response probabilities over the $C_j$ response categories. Similar consistency results to those in Theorem 1 can be established. We leave this extension for future investigation. 
 
Another future direction is to extend the sparse LCA framework to multi-group LCA \citep{clogg1985simultaneous,mccutcheon2002}, an extension of classical LCA that has received wide real-world applications. 
The multi-group LCA concerns data with observed groups (e.g., gender, ethnicity, country) and aims to assess differences between these groups using LCA. The item response probabilities in multi-group LCA can be complex, as they depend on both latent classes and observed groups. A sparse estimate may lead to clearer item response probabilities and thus improve the interpretability of the resulting multi-group LCA model. 

{The proposed procedure is based on a pseudo-likelihood approach. An alternative strategy would be to adopt a full likelihood framework incorporating an $L_0$-type penalty on the differences $\beta_{jk} - \beta_{jk'}$ for $1 \leq k < k' \leq K$, thereby encouraging fusion of item-response probabilities across classes. This idea is in the spirit of the fusion approach for levels of categorical predictors in high-dimensional logistic regression proposed by \cite{Kaufmann2024Simultaneous}. Future research could develop this likelihood-based formulation and compare its performance with the method proposed here. }

Finally, we note that the current methodology and theoretical framework were developed for the low-dimensional regime, in which the sample size $N$ is much larger than the number of items $J$. However, modern applications of latent class analysis often encounter high-dimensional settings in which $J$ is comparable to, or even exceeds, $N$. Because the proposed method and theory may not scale well to such settings, new frameworks are required. To address this high-dimensional latent class analysis problem, future research could adapt the methodologies and theoretical insights developed for high-dimensional latent factor models, such as those in \cite{chen2019joint,chen2020structured} and \cite{chen2022determining}.







 
\bibliographystyle{apalike} 
\bibliography{ref} 

\end{document}